\documentclass[onecolumn]{elsart}%
\usepackage{amssymb}
\usepackage{amsfonts}
\usepackage{amsmath}%
\usepackage{graphicx}
\providecommand{\U}[1]{\protect\rule{.1in}{.1in}}
\begin{document}
\begin{frontmatter}
\title{IS THE FERMI  FIELD CONTACT AND ISOTROPIC?}
\author{E V Rosenfeld}
\address{Ural Division of Russian Academy of Sciences,  Institute of Metal Physics
Kovalevskaya str. 18,  Ekaterinburg 620219, Russia
e-mail: rosenfeld@imp.uran.ru}
\begin{abstract}
It is shown that the contribution to the induction which at an internal point of a spin
density distribution is mathematically described as a local is virtually caused by the
summing-up of the fields created by all elements of this distribution. Therefore, the
proportionality coefficient between this contact (Fermi) field and magnetic moment
density at the point of observation is equal to $8\pi /3$ only for spherically symmetrical
$s$-shells. If the symmetry of spin density distribution lowers, the value of this
coefficient becomes dependent on the spin direction. As a sequence, in low-symmetry
crystals and molecules additional anisotropic contributions to the hyperfine field emerge.
PACS:   76.60.Jx
\end{abstract}
\begin{keyword}
Fermi-contact field anisotropy
\end{keyword}
\end{frontmatter}

\section*{Introduction}

Spin magnetic moments of electrons from proper shells of an isolated atom are
commonly treated to make two contributions to the hyperfine field
$\boldsymbol{B}_{hf}$ on its nucleus\emph{\ }\cite{Bleaney}. The first is the
contact Fermi field produced by $s$-electrons \cite{Fermi}, which in the Gauss
system of units is equal to%

\begin{equation}
\boldsymbol{B}_{F}=-\frac{8\pi}{3}\mu_{B}\left\vert \Psi_{s}\left(  0\right)
\right\vert ^{2}\boldsymbol{\sigma}\text{\textbf{.}}\label{Eq1}%
\end{equation}

The second is the dipole field created by the distribution of density of spin
magnetic moment of electrons from the shells with non-zero orbital moment
$l\neq0$ \cite{Bleaney,White}%
\begin{equation}
\boldsymbol{B}_{dip}\left(  \boldsymbol{\sigma}\right)  =-\mu_{B}%
%TCIMACRO{\dint }%
%BeginExpansion
{\displaystyle\int}
%EndExpansion
\left\vert \Psi_{l}\left(  \boldsymbol{r}\right)  \right\vert ^{^{2}}%
\frac{3\left(  \boldsymbol{r\sigma}\right)  \boldsymbol{r}\mathbf{-}%
r^{2}\boldsymbol{\sigma}}{r^{5}}d\boldsymbol{r}\text{.}\label{Eq2}%
\end{equation}
In these formulas a nucleus is considered to be placed in the point of origin
$r=0$; $\mu_{B}$ is the Bohr magneton; $\boldsymbol{\sigma}$ is the unit
vector in the direction of the spin moment of electron; $\Psi\left(
\boldsymbol{r}\right)  $\ - the electron wave function whose subscript in
$\left(  \ref{Eq1}\right)  $, $\left(  \ref{Eq2}\right)  $ designates only the
orbital moment of a shell.

\ The two fields\emph{\ }$\left(  \ref{Eq1}\right)  $\emph{\ }and\emph{\ }%
$\left(  \ref{Eq2}\right)  $\emph{\ }are created by two parts of the same
distribution of electronic spin moment density. Nevertheless, the contact
Fermi field $\left(  \ref{Eq1}\right)  $ apparently differs from the dipole
field $\left(  \ref{Eq2}\right)  $ in that it is dependent solely on the value
of the wave function at $r=0$ whereas the latter is controlled by the
distribution of spin density all over around the nucleus. And what is more,
the $B_{F}$ value is independent of $\boldsymbol{\sigma}$, while the value of
$B_{dip}$ depends on the $\mathbf{\sigma}$ orientation relative to selected
directions of the\ spin density distribution. That is why it seems appropriate
to formulate the following questions herein.

(i) Why does not the Fermi field depend on the distribution of spin density
all over around the nucleus?

(ii) Why is the multiple in $B_{F}$ equal namely to $8\pi/3$?

(iii) Would the $B_{F}$ value change if the shape of $\Psi_{s}\left(
\boldsymbol{r}\right)  $ changed whereas the $\Psi_{s}\left(  0\right)  $
value does not change?

In what follows the answers for these questions are proposed based in part on
the results related the field acting on a muon at interstice in solids
\cite{Rosenfeld 1}.

\section{The local contribution to magnetic induction}

In nonrelativistic quantum mechanics just as in classical electrodynamics,
magnetic field of any magnetization distribution\footnote{In what follows we
will not tell magnetization distribution from the distribution of spin
magnetic moment density.}%

\begin{equation}
\boldsymbol{M}\left(  \boldsymbol{r}\right)  =-\mu_{B}\left\vert \Psi\left(
\boldsymbol{r}\right)  \right\vert ^{2}\boldsymbol{\sigma}\label{Eq3}%
\end{equation}
is defined by integral $\left(  \ref{Eq2}\right)  $. If the point of
observation $r=0$ lies inside the distribution and the spin density does not
vanish in this point, the under-integral expression diverges. Using standard
techniques of calculating the above improper integral (see, for example,
\cite{White,Tamm,Rosenfeld 1} ) and assuming that the magnetization changes
over the space continuously, one can obtain expression
\begin{equation}
\boldsymbol{B}\left(  0\right)  =4\pi\boldsymbol{M}\left(  0\right)
+\int\frac{\boldsymbol{r}~div\left(  \boldsymbol{M}\left(  \boldsymbol{r}%
\right)  \right)  }{r^{3}}d\boldsymbol{r}\text{\textbf{,}}\label{Eq4}%
\end{equation}
In the essence, this is simply another form of the standard formula
$\boldsymbol{B}=4\pi\boldsymbol{M}\mathbf{+}\boldsymbol{H}$ that defines the
magnetic induction inside the magnetization distribution. The first term
herein can be called a local because it, as well as the Fermi field, is
proportional to the magnetization value at the point of observation, though
with another coefficient. The second term in $\left(  \ref{Eq4}\right)  $
explicitly does not depend on $\boldsymbol{M}\left(  0\right)  $ and is
determined only by the way of changing magnetization over ambient
space\footnote{In the classical electrodynamics, it is the latter contribution
that, together with allowance for the discontinuity of magnetization on the
body surface, determines the demagnetizing field.}.

Implicitly, however, the magnitude of integral contribution to $\left(
\ref{Eq4}\right)  $ strongly depends on the $M\left(  0\right)  $ value. If
$M\left(  0\right)  =0$, as is the case of spin density of electrons from
atomic shells with $l\neq0$ (when this contribution coincides with $\left(
\ref{Eq2}\right)  $),\ $M\left(  r\right)  $ first increases upon moving off a
nucleus and then vanishes at $r\rightarrow\infty$. As is easily seen, the
regions of increasing and decreasing magnetization with $r$ make contributions
of different senses into integral in $\left(  \ref{Eq4}\right)  $. As a
result, the magnitude of hyperfine field is controlled mainly by the distance
from a nucleus to the area where spin density is mostly localized, i.e., by
the value of parameter $\left\langle r^{-3}\right\rangle $. This designation
in \cite{Bleaney,Watson} was used for the radial part of integral $\left(
\ref{Eq2}\right)  $, and with the designation for the angular integral part
taken also after these authors, one can write the dipole contribution in the form%

\begin{equation}
\left(  \boldsymbol{\sigma}\cdot\boldsymbol{B}_{dip}\left(  \boldsymbol{\sigma
}\right)  \right)  =-\mu_{B}\left\langle r^{-3}\right\rangle \ \left\langle
3\cos^{2}\theta-1\right\rangle \label{Eq5}%
\end{equation}

Unlike is the case when $M\left(  0\right)  \neq0$ and magnetization falls
monotonically to zero with increasing $r$. Providing the distribution of
magnetization is isotropic, $\boldsymbol{M}\left(  \boldsymbol{r}\right)
=\boldsymbol{M}\left(  r\right)  $ ($s$-electrons), it vanishes at
$r=r_{0}^{\left(  1\right)  }$ where the first node of the radial $s$-function
is located (for $1s$-electrons $r_{0}^{\left(  1\right)  }\rightarrow\infty$).
The integral in $\left(  \ref{Eq4}\right)  $ over the region $r>r_{0}^{\left(
1\right)  }$ is equal to zero (see below), and in the range $0<r<r_{0}%
^{\left(  1\right)  }$ magnetization only lowers, so that this integral turns
out rather large. Besides, because of the spherical symmetry, it does not
depend on the form of function $M\left(  r\right)  $ and always equals
$-\frac{4\pi}{3}\boldsymbol{M}\left(  0\right)  $. It is the sum of two terms
in $\left(  \ref{Eq4}\right)  $ that results in the Fermi field $\left(
\ref{Eq1}\right)  $.

What would change in the situation described if the part of spin density that
does not vanish at $r\mathbf{\rightarrow}0$\ is not of spherical symmetry? To
answer this question it's enough to distinguish the two parts of the spin
density distribution. The first of them is spherically symmetrical afresh and
creates the same field $8\pi\boldsymbol{M}(0)/3$ at the point of observation.
The second one has lower symmetry and vanishes in this point, and hence its
contribution to the field is described by the formula $\left(  \ref{Eq5}%
\right)  $. So the field at the point of observation now is not equal to Fermi
field and, which is more, its value depends on $\boldsymbol{M}$ direction.

For instance, in the case of the spin density distribution when the surfaces
of constant magnetization have the shape of ellipsoids, a well-known result
\cite{Tamm} is obtained from $\left(  \ref{Eq4}\right)  $\ %

\begin{equation}
\boldsymbol{B}_{hf}\left(  r=0,\mathbf{\sigma}\right)  =4\pi\left[
1-\mathcal{N}\left(  \mathbf{\sigma}\right)  \right]  \boldsymbol{M}\left(
r=0\right)  \text{,}\label{Eq6}%
\end{equation}
where $\mathcal{N}\left(  \mathbf{\sigma}\right)  $ is the demagnetizing
factor of an ellipsoid. In other words, the magnitude of integral in $\left(
\ref{Eq4}\right)  $, still remaining proportional to $M\left(  0\right)  $,
turns out dependent also on the direction of spin moment, which is designated
as previously by symbol $\mathbf{\sigma}$. For the spherically symmetrical
distribution of magnetization this dependence disappears, $\mathcal{N}\left(
\mathbf{\sigma}\right)  \equiv1/3$ so that $4\pi\left(  1-\mathcal{N}\right)
=8\pi/3$ and one comes again to the Fermi field $\left(  \ref{Eq1}\right)  $.

Let us discuss now how local expressions $\left(  \ref{Eq1}\right)  $ and
$\left(  \ref{Eq6}\right)  $ arise from the integral in $\left(
\ref{Eq4}\right)  .$ The matter is that a homogeneously magnetized layer that
lies between two similar ellipsoidlike surfaces (in particular, spherical
layer) does not produce the field inside \cite{Tamm}. Representing the whole
magnetization distribution in the form of an "onion" consisting of such
closely adjoined layers, we can take off any external layers without changing
field in the center \footnote{It is just the reason for the part of spin
density of $s$-electron from the region $r>r_{0}^{\left(  1\right)  }$ to not
contribute to hyperfine field.}. As a result, the field in the center turns
out coinciding with the field of an infinitely small homogeneously magnetized
ellipsoid, and the same conclusion follows from the common simple method of
calculating Fermi field \cite{White} as well.

However, it does not mean that the Fermi field is produced solely by the
infinitesimal central part of the spin density. It means only that for some
spin density distributions of high symmetry (in particular spherical, cubic
and ellipsoid-like symmetry) the sum of contributions to $\boldsymbol{B}_{hf}
$, produced by the outer parts of the distribution, becomes equal to zero.
This is not true for the spin density distributions of lower symmetry, and to
make matters worse, in this case one can not find unambiguously the part of
the spin density that does not vanish in the point of observation, i.e. the
part that should create the \textquotedblleft contact\textquotedblright%
\ field. If, in addition to this, one takes into account the $\boldsymbol{B}%
_{hf}$ anisotropy, which is absent only in a spherically symmetrical case
$\left(  \ref{Eq1}\right)  $, to infer on the "locality" is hardly
pertinent.\newline

\section{Spin density anisotropy and Fermi field in crystals and molecules}

Could, however, one expect that in an atom there is a part of spin density
that would lower monotonically and anisotropically upon moving off the
nucleus? It is impossible of course in an isolated atom, but it may be the
case in a crystal of low-symmetry. Apparently, it is how, for example, the
density of itinerant electrons in crystals with the lattice symmetry lower
than cubic should behave. The simplest argument in favor of this quite evident
statement can be obtained in the approximation of almost free
electrons.\emph{\ }Consider an electronic state\emph{\ }$\left\vert
\boldsymbol{k}\right\rangle =\exp\left(  i\boldsymbol{kr}\right)  $\emph{\ }in
the rhombic $\Gamma_{0}$-type lattice. Let the perturbation potential has the form%

\begin{equation}
\overset{\symbol{94}}{V}\left(  \boldsymbol{r}\right)  =2V\left[  \cos\left(
\frac{2\pi x}{a}\right)  +\cos\left(  \frac{2\pi y}{b}\right)  +\cos\left(
\frac{2\pi z}{c}\right)  \right]  \text{.}\label{Eq7}%
\end{equation}
If the\emph{\ }$V$\emph{\ }value is quite small comparing to the change of
kinetic energy of an electron $\varepsilon\left(  \boldsymbol{k}\right)
=\frac{\hslash^{2}k^{2}}{2m}$, in the lowest order of the perturbation theory,
six states $\left\vert k_{x}+\alpha\frac{2\pi}{a},k_{y},k_{z}\right\rangle $,
$\left\vert k_{x},k_{y}+\alpha\frac{2\pi}{b},k_{z}\right\rangle $, $\left\vert
k_{x},k_{y},k_{z}+\alpha\frac{2\pi}{c}\right\rangle $, $\ \alpha=\pm1$ are
mixed to each state $\left\vert \boldsymbol{k}\right\rangle $ with one and the
same matrix element $V$.\emph{\ }Hence, the charge $\rho\left(  \boldsymbol{r}%
\right)  $ and spin magnetic moment $M\left(  \boldsymbol{r}\right)  $ density
of an electron acquire an inhomogeneous addendum proportional to%

\begin{equation}
-\frac{2mV}{\hslash^{2}}\left[  \frac{\cos\left(  \frac{2\pi x}{a}\right)
}{\left(  \frac{\pi}{a}\right)  ^{2}-\left(  k_{x}\right)  ^{2}}+\frac
{\cos\left(  \frac{2\pi y}{b}\right)  }{\left(  \frac{\pi}{b}\right)
^{2}-\left(  k_{y}\right)  ^{2}}+\frac{\cos\left(  \frac{2\pi z}{c}\right)
}{\left(  \frac{\pi}{c}\right)  ^{2}-\left(  k_{z}\right)  ^{2}}\right]  .
\end{equation}
\bigskip

Hence, the distribution of spin density around the point of origin (the
nucleus \ of one of the atoms constituent \ the lattice) is already
anisotropic. In the vicinity of a nucleus, isolines of spin density turn out
to have the form of ellipsoids elongated the stronger, the greater is the
difference in lattice parameters $a,b$ and $c$.

Similar anisotropy should arise in the opposite limiting case as well, i.e. in
the tight binding approximation, if to allow for the fact that the wave
functions of neighbors overlap the stronger, the closer they are located. To
make it evident, a study of the simplest case of a two-atom molecule is
sufficient. Let us treat an ionized hydrogen molecule $H_{2}^{+}$ in which
atomic nuclei have the coordinates $\boldsymbol{r}_{1}$and $\boldsymbol{r}%
_{2}$, with the vector $\boldsymbol{R}=\boldsymbol{r}_{2}-\boldsymbol{r}_{1}$
being parallel to the $z$ axis. The ground-state wave function is chosen in
the form%

\begin{align}
\Psi\left(  \boldsymbol{r}\right)   & =\frac{1}{Q}\left[  e^{-\left\vert
\boldsymbol{r}-\boldsymbol{r}_{1}\right\vert /a}+e^{-\left\vert \boldsymbol{r}%
-\boldsymbol{r}_{2}\right\vert /a}\right]  ,\label{Eq9}\\
Q  & =\sqrt{2\pi a^{3}\left[  1+\left(  1+\rho+\rho.^{2}/3\right)  \exp\left(
-\rho\right)  \right]  }.\nonumber
\end{align}
Here, $\rho=R/a$, $a=a_{0}/Z$ is the radius of the $1s$ orbit, which at
$R\longrightarrow\infty$ is equal to the Bohr radius $a_{0}$ and decreases
when the atoms come closer \cite{Fermi1}. This effect is taken into account
via introduction of an effective nucleus charge $Z$ whose value is conditioned
at each $R$ by the minimal average energy of the state $\left(  \ref{Eq9}%
\right)  $.

The spin magnetization distribution in this state is the sum of two contributions:%

\begin{align}
\boldsymbol{M}\left(  \boldsymbol{r}\right)   & =\boldsymbol{M}_{1}\left(
\boldsymbol{r}\right)  +\boldsymbol{M}_{2}\left(  \boldsymbol{r}\right)
,\nonumber\\
\boldsymbol{M}_{1}\left(  \boldsymbol{r}\right)   & =-\frac{\mu_{B}%
\mathbf{\sigma}}{Q^{2}}\left(  e^{-2\left\vert \boldsymbol{r}-\boldsymbol{r}%
_{1}\right\vert /a}+e^{-2\left\vert \boldsymbol{r}-\boldsymbol{r}%
_{2}\right\vert /a}\right)  \mathbf{,}\label{Eq10}\\
\boldsymbol{M}_{2}\left(  \boldsymbol{r}\right)   & =-\frac{2\mu
_{B}\mathbf{\sigma}}{Q^{2}}e^{-\left(  \left\vert \boldsymbol{r}%
-\boldsymbol{r}_{1}\right\vert +\left\vert \boldsymbol{r}-\boldsymbol{r}%
_{2}\right\vert \right)  /a}\mathbf{.}\nonumber
\end{align}
\newline The magnetization distribution $\boldsymbol{M}_{1}\left(
\boldsymbol{r}\right)  $ creates at every nucleus the magnetic field induction:%

\begin{align}
\boldsymbol{B}_{1}  & =-\frac{8\pi}{3}\mu_{B}\boldsymbol{\sigma}%
\frac{1+e^{-2\rho}}{Q^{2}}+\boldsymbol{B}_{dip}\mathbf{,}\label{Eq11}\\
\boldsymbol{B}_{dip}  & \mathbf{=}\frac{3\left(  \boldsymbol{\mu R}\right)
\boldsymbol{R}-\boldsymbol{\mu}R^{2}}{R^{5}},\nonumber
\end{align}
This result can easily be obtained after simple reasoning similar to the above
with account for the fact that a homogeneously magnetized sphere creates
around the same field as a point dipole placed in the center of the sphere and
having magnetic moment equal to that of the sphere.

The formula $\left(  \ref{Eq11}\right)  $ looks like the standard expression
for calculation of the field at an atomic nucleus in a crystal, which includes
the Fermi contribution and the dipole fields created by the neighboring
magnetic atoms. However, in this case, the homogeneous and equal to
$\mathbf{\sigma}\exp\left(  -2\rho\right)  /Q^{2}$ part of the spin density
distribution, which is concentrated inside the sphere with a radius $R$ and
the center located at the neighboring site, contributes solely to the Fermi
field. The dipole field is given rise to only by a heterogeneous part of this
distribution so that the magnetic moment entering into $\boldsymbol{B}_{dip}$
$\left(  \ref{Eq11}\right)  $ is equal%

\begin{align}
\boldsymbol{\mu}  & \mathbf{=}\mathbf{-}\frac{4\pi\mu_{B}\boldsymbol{\sigma}%
}{Q^{2}}\int\nolimits_{0}^{R}x^{2}\left(  e^{-2x/a}-e^{-2\rho}\right)
dx\label{Eq12}\\
& =\frac{1}{2}\boldsymbol{\sigma}\mu_{B}\frac{1-\left(  1+2\rho+2\rho
^{2}+4\rho^{3}/3\right)  \exp\left(  -2\rho\right)  }{1+\left(  1+\rho
+\rho^{2}/3\right)  \exp\left(  -\rho\right)  }.\nonumber
\end{align}
As a result, with decreasing $R$, the anisotropic contribution $B_{dip}$
reaches its maximum and vanishes at $R\rightarrow0$, rather then increases
proportionally to $1/R^{3}$.

The second contribution to the magnetization $\boldsymbol{M}_{2}\left(
\boldsymbol{r}\right)  $ arises because of the interference of the atomic
states in $\left(  \ref{Eq9}\right)  $. The surfaces \ $M_{2}\left(
\boldsymbol{r}\right)  =const$ are the ellipsoids of revolution, with the
eccentricity tending to zero at $\left\vert \boldsymbol{r}-\boldsymbol{r}%
_{1}\right\vert +\left\vert \boldsymbol{r}-\boldsymbol{r}_{2}\right\vert
\longrightarrow R$. Again, with the considerations similar to the
above-employed, it is easy to show that the contribution of $\boldsymbol{M}%
_{2}$ into the magnetic field induction has the same value and direction at
the nuclei and at every point of the segment that connects them. Just as the
Fermi field, this contribution is proportional to the magnitude
$\boldsymbol{M}_{2}\left(  \boldsymbol{r}\right)  $, which is constant at
every point of the segment and equal to $-2\mu_{B}\mathbf{\sigma}\exp\left(
-\rho\right)  /Q^{2}$. Therefore, the field induction related to
$\boldsymbol{M}_{2}$ is determined by the formula similar to $\left(
\ref{Eq6}\right)  $:%

\begin{align}
\boldsymbol{B}_{2}\left(  \mathbf{\sigma}\right)   & =-8\pi\mu_{B}%
\mathbf{\sigma}\left[  1-\mathcal{N}\left(  \mathbf{\sigma}\right)  \right]
\exp\left(  -\rho\right)  /Q^{2}\text{,}\nonumber\\
\mathcal{N}\left(  \mathbf{\sigma\parallel}\overset{\symbol{94}}{z}\right)   &
=\frac{1}{2}\mathcal{I}\left(  \rho\right)  -2\frac{\rho+1}{\rho^{2}%
},\label{Eq13}\\
\mathcal{N}\left(  \mathbf{\sigma\perp}\overset{\symbol{94}}{z}\right)   &
=\frac{\left(  1+\rho\right)  ^{2}+1}{2\rho^{2}}-\frac{1}{4}\mathcal{I}\left(
\rho\right)  ,\nonumber\\
\mathcal{I}\left(  \rho\right)   & =\rho\int_{0}^{\infty}e^{-\rho x}x\left(
x+1\right)  \left(  x+2\right)  \ln\left(  \frac{x+2}{x}\right)  dx.\nonumber
\end{align}
%

%TCIMACRO{\FRAME{ftbpFU}{5.2027in}{3.7083in}{0pt}{\Qcb{The dependences on
%$R/a_{0}$ of two contributions into the hyperfine field at nuclei of the
%molecule $H_{2}^{+}$ - $B_{\QTR{bf}{1}}$\ $\left(  \ref{Eq11}\right)  $ and
%$B_{\QTR{bf}{2}}$\ $\left(  \ref{Eq13}\right)  $ for different direction of
%the spin $\QTR{bf}{\sigma}$. In the first insert, the dependence of the
%anisotropic parts of these contributions - $B_{dip}\QTR{bf}{\left(
%\QTR{Bbb}{\sigma}\right)  }$ and $\ B_{Fan}\QTR{bf}{\left(  \QTR{Bbb}{\sigma
%}\right)  =}B_{\QTR{bf}{2}}\QTR{bf}{\left(  \QTR{bf}{\sigma}\right)
%-}16\pi\mu\QTR{bf}{\exp\left(  -\rho\right)  /}\left(  3Q^{2}\right)  $on
%$R/a_{0}$ are shown, while in the second insert, the ratio $B_{Fan}/B_{dip}%
%$\ independent of $\QTR{bf}{\sigma}$.}}{\Qlb{Fig.1}}{graph1.eps}%
%width 5.2027in;  height 3.7083in;  depth 0pt;  original-width 1.3759in;
%original-height 0.9729in;  cropleft "0";  croptop "1";  cropright "1";
%cropbottom "0";  filename 'Figure1.WMF';file-properties "XNPEU";}}}%
%BeginExpansion
\begin{figure}
[ptb]
\begin{center}
\includegraphics[%natheight=0.972900in,
%natwidth=1.375900in,
%height=3.7083in,
%width=5.2027in
]%
{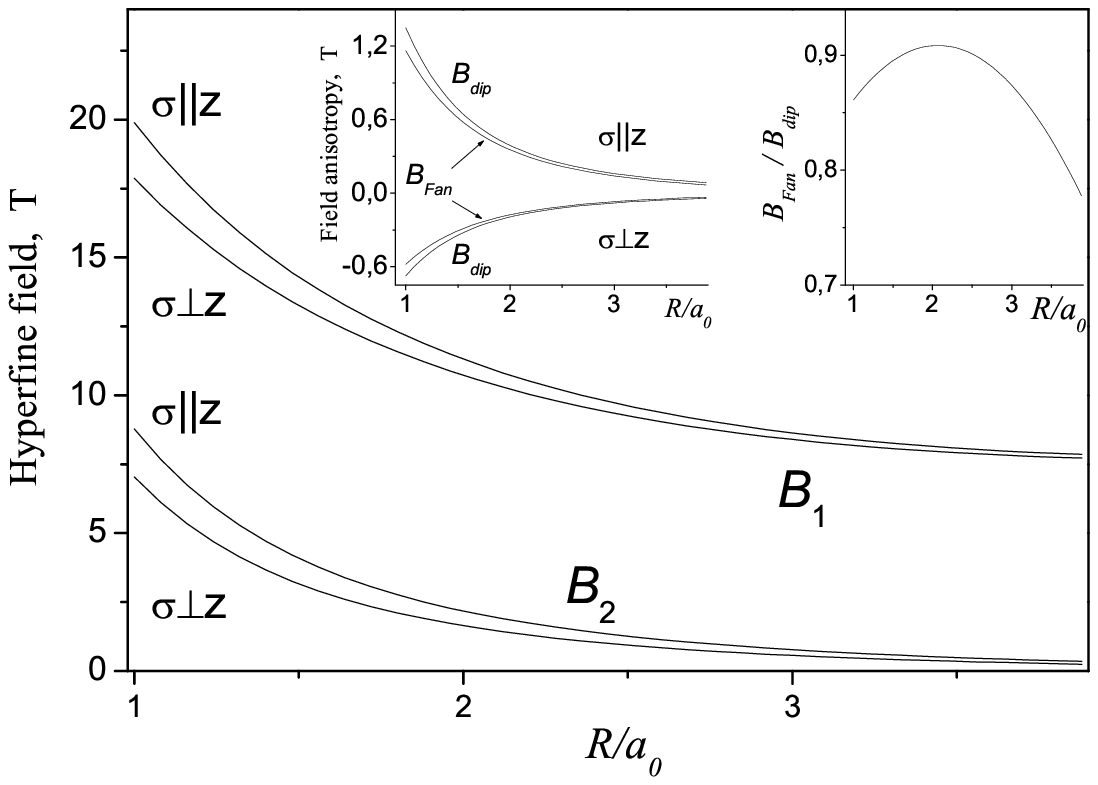}%
\caption{The dependences on $R/a_{0}$ of two contributions into the hyperfine
field at nuclei of the molecule $H_{2}^{+}$ - $B_{\mathbf{1}}$\ $\left(
\ref{Eq11}\right)  $ and $B_{\mathbf{2}}$\ $\left(  \ref{Eq13}\right)  $ for
different direction of the spin $\mathbf{\sigma}$. In the first insert, the
dependence of the anisotropic parts of these contributions - $B_{dip}%
\mathbf{\left(  \mathbb{\sigma}\right)  }$ and $\ B_{Fan}\mathbf{\left(
\mathbb{\sigma}\right)  =}B_{\mathbf{2}}\mathbf{\left(  \mathbf{\sigma
}\right)  -}16\pi\mu\mathbf{\exp\left(  -\rho\right)  /}\left(  3Q^{2}\right)
$on $R/a_{0}$ are shown, while in the second insert, the ratio $B_{Fan}%
/B_{dip}$\ independent of $\mathbf{\sigma}$.}%
\label{Fig.1}%
\end{center}
\end{figure}
%EndExpansion

The dependences of $B_{1}$ and $B_{2}$ on $R/a_{0}$ at $\boldsymbol{\sigma
}\mathbf{\parallel}z$\ and $\boldsymbol{\sigma}\mathbf{\perp}z$ \ are shown in
the Figure \ref{Fig.1} and corresponding dependences of the anisotropic
contributions to $B_{1}$ and $B_{2}$, i.e., $B_{dip}\mathbf{\left(
\boldsymbol{\sigma}\right)  }$ and $B_{Fan}\mathbf{\left(  \boldsymbol{\sigma
}\right)  =}B_{\mathbf{2}}\mathbf{\left(  \boldsymbol{\sigma}\right)
-}\left[  B_{\mathbf{2}}\mathbf{\left(  \boldsymbol{\sigma}\mathbf{\parallel
}\overset{\symbol{94}}{z}\ \right)  +2}B_{\mathbf{2}}\mathbf{\left(
\boldsymbol{\sigma}\mathbf{\perp}\overset{\symbol{94}}{z}\right)  }\right]
/3$, are displayed in the first insert. The total anisotropy value makes up
about 10\% $\left\vert \boldsymbol{B}_{1}+\boldsymbol{B}_{2}\right\vert $, and
the Fermi-field anisotropy turns out only slightly lower than the dipole field
anisotropy, as is seen from the second insert in the same figure.

Thus, in the $H_{2}^{+}$ molecule, with decreasing spacing, not only the
hyperfine field at the nuclei as such increases (chemical shift) but its
anisotropy as well. In this case, about half the anisotropy contribution is
ascribed to the anisotropic part of the Fermi field related to hybridization
of the atomic orbitals. The same effect should evidently arise also in
crystals in which the distances from a magnetic ion to its closest neighbors
in different directions strongly vary, which leads to various overlapping of
their outer $s$-shells.

In general, the value of local contribution to the induction $\left(
\ref{Eq4}\right)  $ is controlled by the form of dependence $div\left(
\boldsymbol{M}\right)  $ on coordinates. If the spin part of the wave function
does not depend on $\boldsymbol{r}$, i.e. $\boldsymbol{M}\left(
\boldsymbol{r}\right)  $ is directed parallel to $z$-axe in any point,
$div\left(  \boldsymbol{M}\right)  $\ has the form%
\begin{equation}
div\left(  \boldsymbol{M}\left(  \boldsymbol{r}\right)  \right)
=\frac{\partial M}{\partial r}\cos\theta-\frac{1}{r}\frac{\partial M}%
{\partial\theta}\sin\theta.\label{Eq14}%
\end{equation}
The first term in this formula having been integrated on $\boldsymbol{r}$ in
$\left(  \ref{Eq4}\right)  $ always gives $-\frac{4\pi}{3}M\left(  0\right)
$. \ The result of the integration of the second term in common case is the
functional of $M\left(  \boldsymbol{r}\right)  $. In particular, if $M\left(
\boldsymbol{r}\right)  =\mathcal{M}\left(  r\right)  \Phi\left(
\theta,\varphi\right)  $, as is the case for atomic states with\emph{\ }%
$l\neq0$\emph{,} the\ \ value of \ this functional depends on $\left\langle
r^{-3}\right\rangle $ just as in $\left(  \ref{Eq5}\right)  $. For the result
of this integration to be, alike the Fermi field $\left(  \ref{Eq6}\right)  $,
proportional solely to $M\left(  0\right)  $, the $M\left(  \boldsymbol{r}%
\right)  $ dependence should have the special form
\begin{equation}
M\left(  \boldsymbol{r}\right)  =\mathcal{M}\left(  r\cdot\Phi\left(
\theta,\varphi\right)  \right) \label{Eq15}%
\end{equation}
where $\mathcal{M}$ and $\Phi$ are any sufficiently smooth functions. Taking
into account that in this case\emph{\ }%
\begin{equation}
\frac{1}{r}\frac{\partial M}{\partial\theta}\equiv\frac{\partial M}{\partial
r}\frac{1}{\Phi\left(  \theta,\varphi\right)  }\frac{\partial\Phi}%
{\partial\theta}\text{,}\label{Eq16}%
\end{equation}
we obtain replacing $\left(  \ref{Eq14}\right)  $\emph{\ }and\emph{\ }$\left(
\ref{Eq16}\right)  $\emph{\ }into\emph{\ }$\left(  \ref{Eq4}\right)  $\emph{:}%
\begin{equation}
\boldsymbol{B}\left(  0\right)  =\boldsymbol{M}\left(  0\right)  \left\{
4\pi-\int\left[  \cos\theta-\frac{\sin\theta}{\Phi\left(  \theta
,\varphi\right)  }\frac{\partial\Phi}{\partial\theta}\right]  \sin\theta
d\mathbf{\theta}d\varphi\right\}  \text{.}\label{Eq17}%
\end{equation}
\emph{\ }The atomic $s$-state is a trivial (the angular dependence is absent)
particular case of spin density distribution\emph{\ }$\left(  \ref{Eq15}%
\right)  $. The other examples of distributions of this type are
ellipsoid-like distribution or a distribution with cubic symmetry.

In order that the spin density distribution for an isolated atom turns out a
function of the type $\left(  \ref{Eq15}\right)  $, its ground state should be
mixed to by the states of the continuous spectrum. Such mixing up actually
arises when an atom is located in an electrical field, in particular, the
crystalline electrical field. However, calculation of the corresponding
contribution to the Fermi field is a serious mathematical problem, which
requires special treatment.

\section{Conclusions}

Returning to the questions posed at the beginning of the paper, we now can
formulate the following conclusions.

(i) The contact contribution to the induction at internal point of any
magnetization distribution virtually results from the summing up of the fields
created by all elements of this distribution.

(ii) The value of the proportionality coefficient between the induction and
magnetic moment density at a point of observation is controlled by the
symmetry of the magnetization distribution. This coefficient is isotropic and
equal to $8\pi/3$ only for distributions with a spherical or cubical symmetry.

(iii) In molecules and crystals of a lower symmetry owing to the hybridization
of the states of neighboring atoms, the symmetry of the $s$-electron density
distribution around the nucleus decreases. This results not only in the
proportionality coefficient between the induction and magnetic-moment density
at the point of observation changes, but starts depend on the moment direction.

It should be underlined also, that the main result of this paper reduces to
the substantiation of a very simple statement: individualization of the Fermi
field from the total hyperfine field becomes empty of meaning for low-symmetry
crystals and molecules. The calculation of the field created by a spin density
distribution of low symmetry should be performed by the formula $\left(
\ref{Eq4}\right)  $ or some equivalent to it. The standard presentation of
hyperfine\emph{\ }field as the sum of the Fermi $\left(  \ref{Eq1}\right)  $,
dipole and Lorentz fields and demagnetizing field of the specimen results in
masking the role of the spin density inhomogeniouties, which may be the source
of some faults.

Allowance for this circumstance may be quite important when analyzing the
results of the hyperfine field measurements. The case in hand may be both the
fields acting on a probe particle (neutron, muon and others) in the
interstices \cite{Rosenfeld 1}, and the field on an atomic nucleus in a
crystal or molecule. In all these cases, with a proper account for the
contribution of collective electrons and/or distortions of the atomic
$s$-shells, additional anisotropic contributions to the Fermi field can
emerge. In particular, one could not exclude that the high (up to $3\div4\quad
T$) anisotropy of the hyperfine field at the nuclei of iron atoms that
constitute the dumb-bells in the compounds $R_{2}Fe_{17}$ \cite{van der
Kraan,Gubbens,Averbuch} is related to this effect rather then to unquenching
of orbital moments.

\textbf{Acknowledgments}

I am grateful to Prof. V.V. Dyakin for useful discussions.

This work has been partly supported by Russian Academy of Sciences, grant
02-02-16440 and Ural Division of RAS, project No 5.

\end{document}